\documentstyle[11pt]{article}

\def\dalemb#1#2{{\vbox{\hrule height .#2pt
        \hbox{\vrule width.#2pt height#1pt \kern#1pt
                \vrule width.#2pt}
        \hrule height.#2pt}}}

\let\a=\alpha    \let\e=\epsilon

\def\nn{\nonumber} \def\bd{\begin{document}} \def\ed{\end{document}}
\def\ds{\documentstyle} \let\fr=\frac \let\bl=\bigl \let\br=\bigr
\let\Br=\Bigr \let\Bl=\Bigl 
\let\bm=\bibitem
\let\na=\nabla
\let\pa=\partial \let\ov=\overline
\def\ie{{\it i.e.\ }} 
\newcommand{\be}{\begin{equation}} 
\newcommand{\ee}{\end{equation}} 
\def\ba{\begin{array}}
\def\ea{\end{array}}
\def\ft#1#2{{\textstyle{{\scriptstyle #1}\over {\scriptstyle #2}}}}
\def\fft#1#2{{#1 \over #2}}
\def\del{\partial}
\def\sst#1{{\scriptscriptstyle #1}}
\def\oneone{\rlap 1\mkern4mu{\rm l}}
\def\e7{E_{7(+7)}}
\def\td{\tilde}
\def\wtd{\widetilde}
\def\im{{\rm i}}
\def\bog{Bogomol'nyi\ }
\newcommand{\ho}[1]{$\, ^{#1}$}
\newcommand{\hoch}[1]{$\, ^{#1}$}
\newcommand{\bea}{\begin{eqnarray}} 
\newcommand{\eea}{\end{eqnarray}} 
\newcommand{\ra}{\rightarrow}
\newcommand{\lra}{\longrightarrow}
\newcommand{\Lra}{\Leftrightarrow}
\newcommand{\ap}{\alpha^\prime}
\newcommand{\bp}{\tilde \beta^\prime}
\newcommand{\tr}{{\rm tr} }
\newcommand{\Tr}{{\rm Tr} } 
\newcommand{\NP}{Nucl. Phys. }
\newcommand{\tamphys}{\it Center for Theoretical Physics,
Texas A\&M University, College Station, Texas 77843}

\newcommand{\auth}{H. L\"u\hoch{\dagger}, C.N. Pope\hoch{\dagger}}

\begin{document}
\begin{flushright}
\hfill{CTP TAMU-46/96}\\
\hfill{IC/96/171}\\
\hfill{hep-th/9609126}\\
\end{flushright}

\vspace{20pt}

\begin{center}
{\large {\bf Black $p$-branes and their Vertical Dimensional Reduction }} 

\vspace{30pt}

\auth

\vspace{15pt}

{\tamphys}
\vspace{20pt}

K.-W. Xu

\vspace{15pt}

{\it International Center for Theoretical Physics, Trieste, Italy }

{\it  and }

{\it Institute of Modern Physics, Nanchang University, Nanchang, China }

\vspace{40pt}

\underline{ABSTRACT}
\end{center}

       We construct multi-center solutions for charged, dilatonic,
non-extremal black holes in $D=4$.  When an infinite array of such
non-extremal black holes are aligned periodically along an axis, the
configuration becomes independent of this coordinate, which can therefore be
used for Kaluza-Klein compactification.  This generalises the vertical
dimensional reduction procedure to include {\it non-extremal} black holes.
We then extend the construction to multi-center non-extremal $(D-4)$-branes
in $D$ dimensions, and discuss their vertical dimensional reduction. 

{\vfill\leftline{}\vfill
\vskip	10pt
\footnoterule
{\footnotesize
	\hoch{\dagger}	Research supported in part by DOE 
Grant DE-FG05-91-ER40633 \vskip	-12pt}  \vskip	10pt
}

\pagebreak
\setcounter{page}{1}

\section{Introduction}

    The supergravity theories that arise as the low-energy limits of string
theory or M-theory admit a multitude of $p$-brane solutions.  In general,
these solutions are characterised by their mass per unit $p$ volume, and the
charge or charges carried by the field strengths that support the solutions.
Solutions can be extremal, in the case where the charges and the mass per
unit $p$-volume saturate a \bog bound, or non-extremal if the mass per unit
$p$-volume exceeds the bound.  There are two basic types of solution, namely
elementary $p$-branes, supported by field strengths of rank $n=p+2$, and
solitonic $p$-branes, supported by field strengths of degree $n=D-p-2$,
where $D$ is the spacetime dimension. Typically, we are interested in
considering solutions in a ``fundamental'' maximal theory such as $D=11$
supergravity, which is the low-energy limit of M-theory, and its various
toroidal dimensional reductions.  A classification of extremal
supersymmetric $p$-branes in M-theory compactified on a torus can be found
in \cite{lpsol}. 

     The various $p$-brane solutions in $D\le 11$ can be represented as
points on a ``brane scan'' whose vertical and horizontal axes are the
spacetime dimension $D$ and the spatial dimension $p$ of the $p$-brane
world-volume.  The same process of Kaluza-Klein dimensional reduction that
is used in order to construct the lower-dimensional toroidally-compactified
supergravities can also be used to perform dimensional reductions of the
$p$-brane solutions themselves:  Since the Kaluza-Klein procedure
corresponds to performing a {\it consistent} truncation of the
higher-dimensional theory, it is necessarily the case that the
lower-dimensional solutions will also be solutions of the higher-dimensional
theory.  There are two types of dimensional reduction that can be carried
out on the $p$-brane solutions. The more straightforward one involves a
simultaneous reduction of the spacetime dimension $D$ and the spatial
$p$-volume, from $(D,p)$ to $(D-1,p-1)$; this is known as ``diagonal
dimensional reduction'' \cite{dhis,lpss1}.  It is achieved by choosing
one of the spatial world-volume coordinates as the compactification
coordinate.  The second type of dimensional reduction corresponds to a
vertical descent on the brane scan, from $(D,p)$ to $(D-1,p)$, implying that
one of the directions in the space transverse to the $p$-brane world-volume
is chosen as the compactification coordinate.  This requires that one first
construct an appropriate configuration of $p$-branes in $D$ dimensions that
has the necessary $U(1)$ isometry along the chosen direction.  It is not
{\it a priori} obvious that this should be possible, in general.  However,
it is straightforward to construct such configurations in the case of
extremal $p$-branes, since these satisfy a no-force condition which means
that two or more $p$-branes can sit in neutral equilibrium, and thus multi
$p$-brane solutions exist \cite{cg}.  By taking a limit corresponding to an
infinite continuum of $p$-branes arrayed along a line, the required
$U(1)$-invariant configuration can be constructed \cite{k,ghl,ht,lps} 

     In this paper, we shall investigate the dimensional-reduction
procedures for non-extremal $p$-branes.  In fact the process of diagonal
reduction is the same as in the extremal case, since the non-extremal 
$p$-branes also have translational invariance in the spatial world-volume
directions.  The more interesting problem is to see whether one can also
describe an analogue of vertical dimensional reduction for non-extremal
$p$-branes.  There certainly exists an algorithm for constructing a
non-extremal $p$-brane at the point $(D-1,p)$ from one at $(D,p)$ on the
brane scan.  It has been shown that there is a universal prescription for
``blackening'' any extremal $p$-brane, to give an associated non-extremal
one \cite{dlp}.  Thus an algorithm, albeit inelegant, for performing the
vertical reduction is to start with the general non-extremal $p$-brane in
$D$ dimensions, and then take its extremal limit, from which an extremal
$p$-brane in $D-1$ dimensions can be obtained by the standard
vertical-reduction procedure described above.  Finally, one can then invoke
the blackening prescription to construct the non-extremal $p$-brane in $D-1$
dimensions.  However, unlike the usual vertical dimensional reduction for
extremal $p$-branes, this procedure does not give any physical
interpretation of the $(D-1)$-dimensional $p$-brane as a superposition of
$D$-dimensional solutions. 

    At first sight, one might think that there is no possibility of
superposing non-extremal $p$-branes, owing to the fact that they do not
satisfy a no-force condition.  Indeed, it is clearly true that one cannot
find well-behaved static solutions describing a finite number of black
$p$-branes located at different points in the transverse space.  However, we
do not require such general kinds of multi $p$-brane solutions for the
purposes of constructing a configuration with a $U(1)$ invariance in the
transverse space.  Rather, we require only that there should exist static
solutions corresponding to an infinite number of $p$-branes, periodically
arrayed along a line.  In such an array, the fact that there is a
non-vanishing force between any pair of $p$-branes is immaterial, since the
net force on each $p$-brane will still be zero.  The configuration is in
equilibrium, although of course it is highly unstable.  For example, one can
have an infinite static periodic array of $D=4$ Schwarzschild black holes
aligned along an axis.    In fact the instability problem is overcome in the
Kaluza-Klein reduction, since the extra coordinate $z$ is compactified on a
circle.  Thus there is a stable configuration in which the $p$-branes are
separated by precisely the circumference of the compactified dimension. 
Viewed from distances for which the coordinates orthogonal to $z$ are large
compared with this circumference, the fields will be effectively independent
of $z$, and hence $z$ can be used as the compactification coordinate for the
Kaluza-Klein reduction, giving rise to a non-extremal $p$-brane in $D-1$
dimensions. 

    In section 2, we obtain the equations of motion for axially symmetric 
$p$-branes in an arbitrary dimension $D$.  We then construct multi-center
non-extremal $D=4$ black hole solutions in section 3, and show how they may 
be used for vertical dimensional reduction of non-extremal black holes.
In section 4, we generalise the construction to non-extremal $(D-4)$-branes
in arbitrary dimension $D$.
     
\section{Equations of motion for axially symmetric $p$-branes}

    We are interested in describing multi-center non-extremal $p$-branes
in which the centers lie along a single axis in the transverse space.
Metrics with the required axial symmetry can be parameterised in the 
following way:
\be
ds^2=-e^{2U}dt^2+e^{2A}dx^idx^i+e^{2V}(dz^2+dr^2)+e^{2B}r^2d\Omega^2
\ ,\label{metricans}
\ee
where $(t,x^i),\,\, i=1,\ldots,p$, are the coordinates of the $p$-brane
world-volume. The remaining coordinates of the $D$ dimensional spacetime, 
\ie those in the transverse space, are $r$, $z$ and the coordinates on a
$\td d$-dimensional unit sphere, whose metric is $d\Omega^2$, with $\td
d=D-p-3$. The functions $U$, $A$, $V$ and $B$ depend on the coordinates
$r$ and $z$ only.  We find that the Ricci tensor for the metric
(\ref{metricans}) is given by 
\bea
R_{00}&=&e^{2U-2V}\Big(U''+\ddot U+U'^2 + \dot U^2 + p(U'A'+\dot U \dot A)+
\td d(U'B' +\dot U \dot B) + \fft{\td d}{r}U'\Big)\ ,\nonumber\\
R_{rr}&=&-\Big(U'' - U'V' + U'^2 +\dot U \dot V +\ddot V + V'' +
p(A'' - V' A' + A'^2 + \dot V \dot A) \nonumber\\
&&+ \td d (B''-V'B'-\fft{1}{r}V' +\dot V \dot B + B'^2 +\fft{2}{r}B')\Big)\ ,
\nonumber\\
R_{zz}&=&-\Big(\ddot U -\dot U \dot V + {\dot U}^2 + U' V' + \ddot V
+V'' +p(\ddot A - \dot V \dot A +{\dot A}^2 + V' A' ) \nonumber\\
&&+ \td d (\ddot B -\dot V \dot B + V' B' + {\dot B}^2 +\fft{1}{r} V')\Big)\ ,
\nonumber\\
R_{rz}&=&\Big( -\dot U' +\dot V U' -\dot U U' +\dot U V' +
p(-\dot A' +\dot V A' -\dot A A' + V' \dot A ) \label{ricci}\\
&&+ \td d (-\dot B' +\dot V B' +\fft{1}{r} \dot V -\fft{1}{r} \dot B -\dot B B'
+V' \dot B )\Big)\ ,\nonumber\\
R_{ab}&=&-e^{2B-2V}\Big(B'' +\ddot B +U' B'+\dot U \dot B +\fft{1}{r} U' +
p(A'B' +\dot A \dot B + \fft{1}{r} A') \nonumber\\
&&+ \td d (B'^2 +{\dot B}^2 +\fft{2}{r} B')\Big)\, r^2\, \bar g_{ab} + 
(\td d-1)(1-e^{2B-2V})\bar g_{ab}\ ,\nonumber\\
R_{ij}&=&-e^{2A-2V}\Big(A'' +\ddot A +U' A' +\dot U \dot A +p(A'^2 +{\dot A}^2)
+\td d (A'B' +\dot A \dot B +\fft{1}{r} A')\Big)\, \delta_{ij}\ ,\nn
\eea
where the primes and dots denote derivatives with respect to $r$ and $z$
respectively, $\bar g_{ab}$ is the metric on the unit $\td d$-sphere, and
the components are referred to a coordinate frame. 

    Let us consider axially-symmetric solutions to the theory described by 
the Lagrangian
\be
e^{-1} {\cal L} =  R -\ft12 (\del \phi)^2 -\fft1{2 n!}\, e^{-a\phi}\, F_n^2\ ,
\label{boslag}
\ee
where $F_n$ is an $n$-rank field strength.  The constant $a$ in the dilaton 
prefactor can be parameterised as
\be
a^2 = \Delta - \fft{2(n-1)(D-n-1)}{D-2}\ ,\label{avalue}
\ee 
where the constant $\Delta$ is preserved under dimensional reduction 
\cite{lpss1}.  (For supersymmetric $p$-branes in M-theory compactified on a
torus, the values of $\Delta$ are $4/N$ where $N$ is an integer $1\le N \le
N_c$, and $N_c$ depends on $D$ and $p$ \cite{lpsol}.  Non-supersymmetric
$(D-3)$-branes with $\Delta = 24/(N(N+1)(N+2))$ involving $N$ 1-form field
strengths were constructed in \cite{lptoda}.  Their equations of motion
reduce to the $SL(N+1,R)$ Toda equations.  Further non-supersymmetric
$p$-branes with other values of $\Delta$ constructed in \cite{lpsol} however
cannot be embedded into M-theory owing to the complications of the
Chern-Simons modifications to the field strengths.) We shall concentrate on
the case where $F_n$ carries an electric charge, and thus the solutions will
describe elementary $p$-branes with $p=n-2$.  (The generalisation to
solitonic $p$-branes that carry magnetic charges is straightforward.)   The
potential for $F_n$ takes the form $A_{0i_1\cdots i_p}= \gamma
\epsilon_{i_1\cdots i_p}$, where $\gamma$ is a function of $r$ and $z$. Thus
the equations of motion will be 
\be
\Box\phi = -\ft12 a \, S^2\ ,\qquad R_{MN} = \ft12\del_M\phi\, \del_N\phi 
+S_{MN}\ ,\qquad \del_{M_1}(\sqrt{-g}\, e^{-a\phi}\, F^{M_1\cdots M_n}) =0
\ ,\label{eom}
\ee
where 
\bea
S_{00} = \fft{\td d}{2(D-2)} S^2 e^{2U}({\dot\gamma}^2+{\gamma'}^2)\ , &&
S_{rr} = \fft{1}{2(D-2)} S^2 e^{2V} ( d {\dot\gamma}^2 -{\td d} {\gamma'}^2
)\ , \nonumber \\ 
S_{zz} = \fft{1}{2(D-2)} S^2 e^{2V} ( -{\td d} {\dot\gamma}^2 + d
{\gamma'}^2)\ , &&
S_{rz} = -\ft{1}{2} S^2 e^{2V} {\dot\gamma} \gamma' \ , \label{sab} \\
S_{ab} = \fft{d}{2(D-2)} S^2 ( {\dot\gamma}^2 + {\gamma'}^2 ) e^{2B} r^2
\bar g_{ab} \ , &&
S_{ij} = -\fft{{\td d}}{2(D-2)} S^2 e^{2A}(\dot\gamma^2 + \gamma'^2)
{\delta}_{ij} \ , \nn 
\eea
$S^2 = e^{-2pA -2V -a\phi-2U} $ and $d=p+1$.

\section{$D=4$ black holes and their dimensional reduction} 

\subsection{Single-center black holes}

  Let us first consider black hole solutions in $D=4$, whose charge is 
carried by a 2-form field strength. By appropriate choice of coordinates,
and by making use of the field equations, we may set $B=-U$. Defining also
$V=K-U$, we find that equations of motion (\ref{eom}) for the metric
\be
ds^2 = -e^{2U}\, dt^2 + e^{2K-2U}\, (dr^2 + dz^2) + e^{-2U}\, r^2\, 
d\theta^2\label{d4metric}
\ee
can be reduced to
\bea
{\nabla}^2 U = \ft{1}{4} e^{-a\phi -2 U} ({\dot\gamma}^2 +{\gamma'}^2) \ ,
&& {\nabla}^2 K - \ft{2}{r} K' = \ft{1}{2} e^{-a\phi - 2 U} {\gamma'}^2 -
2 U'^2 -\ft{1}{2} {\phi'}^2 \ , \nonumber \\
{\nabla}^2 K = \ft{1}{2} e^{-a\phi -2 U} {\dot\gamma}^2 - 2 {\dot U}^2 -
\ft{1}{2} {\dot\phi}^2 \ ,
&& \ft{1}{r} {\dot K} = -\ft{1}{2} e^{-a\phi -2 U} {\dot\gamma} {\gamma'} 
+ 2{\dot U} U' + \ft{1}{2} {\dot\phi} {\phi}' \ , \label{d4eom} \\
{\nabla}^2 \phi = \ft{1}{2} a e^{-a\phi - 2 U} ( {\dot\gamma}^2 
+{\gamma'}^2 ) \ ,
&& {\nabla}^2 {\gamma} = (a \phi' + 2 U') \gamma' + (a {\dot\phi} +
2 {\dot U} ) {\dot\gamma} \, \nn
\eea
where ${\nabla}^2 = {d^2 \over dr^2} + \ft{1}{r} {d \over dr} +
{d^2 \over dz^2} $ is the Laplacian for axially-symmetric functions in
cylindrical polar coordinates. 

     We shall now discuss three cases, with increasing generality, beginning 
with the pure Einstein equation, with $\phi=0$ and $\gamma=0$.  The 
equations (\ref{d4eom}) then reduce to
\be
\nabla^2U=0\ ,\qquad K'=r({U'}^2 -\dot U^2)\ ,\qquad \dot K =2r U'\dot U\ ,
\label{ricflat}
\ee
thus giving a Ricci-flat axially-symmetric metric for any harmonic function 
$U$.  The solution for $K$ then follows by quadratures.
A single Schwarzschild black hole is given by taking $U$ to be the 
Newtonian potential for a rod of mass $M$ and length $2M$ \cite{ik}, \ie 
\be
U = \ft12 \log\fft{\sigma +\td\sigma -2M}{\sigma+\td\sigma +2M} \ ,
\label{rod}
\ee
where $\sigma =\sqrt{r^2+(z-M)^2}$ and $\td \sigma = \sqrt{r^2 +
(z+M)^2}$.  The solution for $K$ is 
\be
K= \ft12 \log\fft{(\sigma+\td\sigma -2M)(\sigma+\td\sigma +2M)}{4\sigma
\td\sigma}\ .\label{ksol1}
\ee
Now we shall show that this is related to the standard Schwarzschild 
solution in isotropic coordinates, \ie
\be
ds^2= - \fft{(1-\fft{M}{2R})^2}{(1+\fft{M}{2R})^2}\, dt^2 +
(1+\fft{M}{2R})^4 (d\rho^2 + dy^2 + \rho^2 d \theta^2) \ ,\label{sch}
\ee
where $R\equiv \sqrt{\rho^2 + y^2}$.  To do this, we note that (\ref{sch}) 
is of the general form (\ref{metricans}), but with $B\ne -U$.  As discussed 
in \cite{s}, setting $B=-U$ depends firstly upon having a field 
equation for which the $R_{00}$ and $R_{\theta\theta}$ components of the 
Ricci tensor are proportional, and secondly upon performing a holomorphic
coordinate transformation from $\eta\equiv r + \im z$ to new variables
$\xi\equiv \rho+ \im y$.  Comparing the coefficients of $d\theta^2$ in
(\ref{d4metric}) and (\ref{sch}), we see that we must have $\Re(\eta)=
\Re(\xi)(1-m^2/(4\bar\xi\xi))$, and hence we deduce that the required
holomorphic transformation is given by 
\be
\eta=\xi -\fft{m^2}{4\xi}\ .\label{trans}
\ee
It is now straightforward to verify that this indeed transforms the metric 
(\ref{d4metric}), with $U$ and $K$ given by (\ref{rod}) and (\ref{ksol1}),
into the standard isotropic Schwarzschild form (\ref{sch}).

    Now let us consider the pure Einstein-Maxwell case, where $\gamma$ is 
non-zero, but $a=0$ and hence $\phi=0$.  We find that the equations of
motion (\ref{d4eom}) can be solved by making the ansatz 
\be
e^{-U} = e^{- {\wtd U} } - c^2 e^{\wtd U} \ , \qquad
\gamma = 2 c e^{ 2 {\wtd U} } \Big( 1 - c^2 e^{ 2 {\wtd U} } \Big)^{-1}
\ , \label{einmax}
\ee
where $c$ is an arbitrary constant and $\wtd U$ satisfies $\nabla^2 \wtd
U=0$.  Substituting into (\ref{d4eom}), we find that all the equations are
then satisfied if 
\be
K' = r\, ( {\wtd U}'^2 - {\dot{\wtd U}}^2 ) \ , \qquad
{\dot K} = 2r\, {\wtd U}'\,  {\dot{\wtd U}} \ . \label{ksol2}
\ee
(Our solutions in this case are in agreement with \cite{g}, after 
correcting some coefficients and exponents.) The solution for a single
Reissner-Nordstr{\o}m black hole is given by taking the harmonic function
$\wtd U$ to be the Newtonian potential for a rod of mass $\ft12 k$ and
length $k$, implying that $\wtd U$ and $K$ are given by 
\bea
\wtd U &=& \ft12 \log\fft{\sigma +\td\sigma -k}{\sigma+\td\sigma +k} 
\ ,\nonumber\\
K&=& \ft12 \log\fft{(\sigma+\td\sigma -k)(\sigma+\td\sigma +k)}{4\sigma
\td\sigma}\ .\label{uk}
\eea
where $\sigma = \sqrt{r^2 + (z-k/2)^2}$ and $\td \sigma =
\sqrt{r^2 + (z+ k/2)^2}$.  The metric can be re-expressed in terms of the
standard isotropic coordinates $(\hat t,\rho,y,\theta)$ by performing the
transformations 
\be
\eta=\fft{1}{1-c^2}\, (\xi -\fft{\hat k^2}{16\xi}) \ ,\qquad
t=(1-c^2)\, \hat t\ ,\label{redef}
\ee
where $\xi=\rho+\im y$ and $\hat k = (1-c^2) k$, giving
\bea
ds^2&=& - \Big(1 + \fft{\hat k R}{(R+\ft14 \hat k)^2} \, \sinh^2\mu\Big)^{-2}\, 
\Big(\fft{R-\ft14 \hat k}{R+\ft14 \hat k}\Big)^2\, d\hat t^2 \nonumber\\
&& + \Big(1 + \fft{\hat k R}{(R+\ft14 \hat k)^2} \, \sinh^2\mu\Big)^2 \,
(1+\fft{\hat k}{4R})^4\, (d\rho^2 + dy^2 + \rho^2 d\theta^2)\ ,\label{rn}
\eea
where $c=\tanh\mu$, and again $R\equiv\sqrt{\rho^2+y^2}$.  (It is necessary
to rescale the time coordinate, as in (\ref{redef}), because the function
$e^{-U}$ given in (\ref{einmax}) tends to $(1-c^2)$ rather than 1 at
infinity.)   Equation (\ref{rn}) is the standard Reissner-Nordstr{\o}m
metric in isotropic coordinates, with mass $M$ and charge $Q$ given in terms
of the parameters $\hat k$ and $\mu$ by 
\be
M= \hat k\sinh^2\mu +\ft12 \hat k\ ,\qquad Q= \ft14 \hat k 
\sinh 2\mu \ . \label{mc}
\ee
The extremal limit is obtained by taking $\hat k\rightarrow 0$ at the same time 
as sending $\mu\rightarrow\infty$, while keeping $Q$ finite, implying that 
$Q=\ft12 M$.  This corresponds to setting $c\rightarrow 1$ in
(\ref{einmax}).  The description in the form (\ref{d4metric}) becomes
degenerate in this limit, since the length and mass of the Newtonian rod
become zero.  However, the rescalings (\ref{redef}) also become singular,
and the net result is that the metric (\ref{rn}) remains well-behaved in the
extremal limit. The previous pure Einstein case is recovered if $\mu$ is
instead sent to zero, implying that $Q=0$ and $c=0$. 

     Finally, let us consider the case of Einstein-Maxwell-Dilaton black 
holes.  We find that the equations of motion (\ref{d4eom}) can be solved by 
making the ans\"atze
\bea
\phi &=& 2a(U -\wtd U)\ ,
\qquad e^{-\Delta U} = (e^{-\wtd U} - c^2\, e^{\wtd U}) e^{-a^2 \wtd U}\ ,
\nonumber\\
\gamma &=& 2c\, e^{2\wtd U} \Big(1 - c^2\, e^{2\wtd U} \Big)^{-1}
\ ,\label{phians}
\eea
where, as in the pure Einstein-Maxwell case, $c$ is an arbitrary constant 
and $\wtd U$ satisfies $\nabla^2 \wtd U=0$.  Substituting the ans\"atze into
(\ref{d4eom}), we find that all the equations are satisfied provided that 
the function $K$ satisfies (\ref{ksol2}).   The solution for a single
dilatonic black hole for generic coupling $a$ is given by again taking the
harmonic function $\wtd U$ to be the Newtonian potential for a rod of mass
$\ft12k$ and length $k$.  After performing the coordinate transformations
\be
\eta=(1-c^2)^{-\ft1{\Delta}}\, (\xi-\fft{\hat k^2}{16\xi})\ ,\qquad
t=(1-c^2)^{\ft{1}{\Delta}}\, \hat t\ ,
\ee
where $\hat k=(1-c^2)^{1/\Delta} k$, and writing $c=\tanh \mu$, we find that
the metric takes the standard isotropic form 
\bea
ds^2&=& - \Big(1 + \fft{\hat k R}{(R+\ft14 \hat k)^2} \, 
\sinh^2\mu\Big)^{-\ft2{\Delta}}\, 
\Big(\fft{R-\ft14 \hat k}{R+\ft14 \hat k}\Big)^2\, d\hat t^2 \nonumber\\
&& + \Big(1 + \fft{\hat k R}{(R+\ft14 \hat k)^2} \,
\sinh^2\mu\Big)^{\ft2{\Delta}} \, (1+\fft{\hat k}{4R})^4\, (d\rho^2 + dy^2 +
\rho^2 d\theta^2)\ .\label{dbh} 
\eea
The mass $M$ and charge $Q$ are given by
\be
M = \fft{\hat k}{\Delta} \sinh^2 \mu + \ft12 \hat k\ ,\qquad
Q= \fft{\hat k}{4\sqrt\Delta} \sinh 2\mu\ .
\ee
Again, the extremal limit is obtained by taking $\hat k \rightarrow 0$, 
$\mu\rightarrow \infty$, while keeping $Q$ finite, implying that 
$Q=\sqrt{\Delta} M/2$.

\subsection{Vertical dimensional reduction of black holes}

    The vertical dimensional reduction of a $p$-brane solution 
requires that the Kaluza-Klein compactification coordinate should lie in the 
space transverse to the world-volume of the extended object.  In order to 
carry out the reduction, it is necessary that the higher-dimensional
solution be independent of the chosen compactification coordinate.  In the 
case of extremal $p$-branes, this can be achieved by exploiting the fact
that there is a zero-force condition between such objects, allowing 
arbitrary multi-center solutions to be constructed. Mathematically,
this can be done because the equations of motion reduce to a Laplace
equation in the transverse space, whose harmonic-function solutions can be 
superposed.  Thus one can choose a configuration with an infinite line of
$p$-branes along an axis, which implies in the continuum limit that the the
solution is independent of the coordinate along the axis. 

      As we saw in the previous section, the equations of motion for an 
axially-symmetric non-extremal black-hole configuration can also be cast in
a form where the solutions are given in terms of an arbitrary solution of 
Laplace's equation.  Thus again we can superpose solutions, to describe 
multi-black-hole configurations.  We shall discuss the general case of 
dilatonic black holes, since the $a=0$ black holes and the uncharged black
holes are merely special cases.  Specifically, a solution in which $\wtd U$
is taken to be the Newtonian potential for a set of rods of mass $\ft12 k_n$
and length $k_n$ aligned along the $z$ axis will describe a line of charged, 
dilatonic black holes: 
\bea
\td U&=& \ft12 \sum_{n=1}^N \log
\fft{\sigma_n +\td \sigma_n -k_n}{\sigma_n +\td \sigma_n +k_n}\ ,
\label{multirod}\\
K &=& \ft14 \sum_{m,n=1}^N \log\fft{[\sigma_m \td\sigma_n +
(z-z_m -\ft12 k_m)(z-z_n+ \ft12 k_n) + r^2]}{[\sigma_m \sigma_n +
(z-z_m -\ft12 k_m)(z-z_n- \ft12 k_n) + r^2]}\label{multisol}\\
&&+\ft14 \sum_{m,n=1}^N \log\fft{[\td\sigma_m \sigma_n +
(z-z_m +\ft12 k_m)(z-z_n- \ft12 k_n) + r^2]}{[\td\sigma_m \td\sigma_n +
(z-z_m +\ft12 k_m)(z-z_n+ \ft12 k_n) + r^2]}\ ,\nonumber
\eea
where $\sigma_n^2 = r^2 + (z-z_n-\ft12 k_n)^2$ and $\td \sigma_n^2 =
r^2 +(z-z_n+\ft12 k_n)^2$. (Multi-center Schwarzschild solutions were 
obtained in \cite{ik}, corresponding to (\ref{multirod}) and (\ref{multisol})
with $k_n=2M_n$, and $U$ equal to $\wtd U$ rather than the expression given 
in (\ref{phians}).)  This describes a system of $N$ non-extremal
black holes, which remain in equilibrium because of the occurrence of
conical singularities along the $z$ axis.  These singularities correspond to
the existence of (unphysical) ``struts'' that hold the black holes in place
\cite{g2,hs}.  If, however, we take all the constants $k_n$ to be equal, and
take an infinite sum over equally-spaced black holes lying at points $z_n=n
b$ along the entire $z$ axis, the conical singularities disappear \cite{m}. 
In the limit when the separation goes to zero, the resulting solution
(\ref{multisol}) becomes independent of $z$.  For small $k=k_n$, we have
$U\sim -\ft12 k (r^2 + (z-nb)^2)^{-1/2}+ O(k^3/r^3)$, and thus in the limit
of small $b$, the sum giving $\wtd U$ in (\ref{multirod}) can be replaced by
an integral: 
\be
\wtd U \sim -\fft{k}{2b} \int_L^L \fft{dz'}{\sqrt{r^2 + z'^2}}\ ,
\ee
in the limit $L\rightarrow\infty$.  Subtracting out the divergent constant 
$-k(\log2 + \log L)/b$, this gives the $z$-independent result \cite{m}
\be
\wtd U = \fft{k}{b} \log r\ .\label{zindep}
\ee
Similarly, one finds that $K$ is given by
\be
K = \fft{k^2}{b^2}\, \log r \ .\label{kzindep}
\ee
One can of course directly verify that these expressions for $\wtd U$ and 
$K$ satisfy the equations of motion (\ref{d4eom}).  Since the associated 
metric and fields are all $z$-independent, we can now perform a dimensional 
reduction with $z$ as the compactification coordinate, giving rise to a 
solution in $D=3$ of the dimensionally reduced theory, which is
obtained from (\ref{boslag}), with $D=4$ and $n=2$, by the standard
Kaluza-Klein reduction procedure.  A detailed discussion of this procedure
may be found, for example, in \cite{lpss1}.  From the formulae given there,
we find that the relevant part of the $D=3$ Lagrangian, namely the part
involving the fields that participate in our solution, is given by 
\be
e^{-1}{\cal L} = R - \ft12 (\del\phi)^2 -\ft12(\del\varphi)^2
-\ft14 e^{-\varphi-a\phi}\, F_2^2\ ,\label{d3lag}
\ee
where $\varphi$ is the Kaluza-Klein scalar coming from the dimensional
reduction of the metric, \ie $ds_4^2 = e^{\varphi}\, ds_3^2 + e^{-\varphi}
dz^2$.  A ``standard'' black hole solution in $D=3$ would be one where only 
the combination of scalars $(-\varphi-a\phi)$, occurring in the exponential 
prefactor of the field strength $F_2$ that supports the solution, is 
non-zero.  In other words, the orthogonal combination should vanish, \ie
$a\varphi-\phi=0$.  Since our solution in $D=4$ has $\phi=2a(U-\wtd U)$, it
follows that $\varphi=2U-2\wtd U$, and hence we should have
\be
ds_4^2 = e^{2U-2\wtd U}\, ds_3^2 + e^{2\wtd U-2U}\, dz^2 \ .
\ee
Comparing this with the $D=4$ solution, whose metric takes the form 
(\ref{d4metric}), we see that the $D=3$ solution will have the above 
single-scalar structure if $K=\wtd U$.  From (\ref{zindep}) and 
(\ref{kzindep}), this will be the case if the parameter $k$ setting the 
scale size of the rods, and the parameter $b$ determining the spacing 
between the rods, satisfy $k=b$.

    It is interesting to note that since $k$ is the length of 
each rod, and $b$ is the period of the array, the condition $k=b$ implies that
the rods are joined end to end, effectively describing a single rod of 
length $L$ and mass $\ft12 L$ in the limit $L\rightarrow \infty$.  In 
other words, the $D=4$ multi-black-hole solution becomes a single black hole 
with $k=L\rightarrow \infty$ in this case.  If $r$ is large compared with 
$z$, the solution is effectively independent of $z$, and thus one can reduce 
to $D=3$ with $z$ as the Kaluza-Klein compactification coordinate. (This is
rather different from the situation in the extremal limit; in that case, the
lengths and masses of the individual rods are zero, and the sum over an
infinite array does not degenerate to a single rod of infinite length.) 

     If $k$ and $b$ are not equal, the dimensional reduction of the $D=4$
array of black holes will of course still yield a 3-dimensional solution of
the equations following from (\ref{d3lag}), but now with the orthogonal
combination $a\varphi -\phi$ of scalar fields active also.  Such a solution
lies outside the class of $p$-brane solitons that are normally discussed;
we shall examine such solutions in more detail in the next section.

\section{Reductions of higher-dimensional black $(D-4)$-branes}

      The equations of motion (\ref{eom}) for general black $p$-branes in 
$D$ dimensions become rather difficult to solve in the axially symmetric 
coordinates, owing to the presence final term involving $(\td
d-1)(1-e^{2B-2V})$ in $R_{ab}$ given in (\ref{ricci}).   This term vanishes 
if $\td d=1$, as it did in the case of 4-dimensional black holes discussed in 
section 3.  The simplest generalisation of these 4-dimensional results is 
therefore to consider $(D-4)$-branes, which have $\td d =1$ also.  They will
arise as solutions of the equations of motion following from (\ref{boslag})
with $n=D-2$.  The required solutions can be obtained by directly solving
the equations of motion (\ref{eom}), with Ricci tensor given by
(\ref{ricci}). However, in practice it is easier to obtain the solutions by
diagonal Kaluza-Klein oxidation of the $D=4$ black hole solutions.  The
ascent to $D$ dimensions can be achieved by recursively applying the inverse
of the one-step Kaluza-Klein reduction procedure. 

     The one step reduction of the metric from $(\ell + 1)$ to $\ell$ 
dimensions takes the form
\be
ds^2_{\ell +1} = e^{2 \a_\ell \varphi_\ell} ds_{\ell}^2 +
                  e^{-2(\ell -2) \a_\ell\varphi_{\ell}} dx_{5-\ell}^2
\ ,\label{kk} \ee 
where $\a_{\ell}^{-2} = 2 (\ell -1) (\ell -2)$. (We have omitted the 
Kaluza-Klein vector potential since it is not involved in the solutions that 
we are discussing.)   The kinetic term for the field strength $F_{\ell-1}$ 
in $(\ell+1)$ dimensions, \ie $e^{-a_{\ell 
+1} \phi_{\ell +1} } F_{\ell-1}^2$, reduces to the kinetic term
$e^{-\a_{\ell +1} \phi_{\ell +1} + 2 \a_\ell \varphi_\ell} F_{\ell -2}^2$ in
$\ell$ dimensions for the relevant field strength $F_{\ell-2}$.  We may 
define $-a_{\ell+1}\, \phi_{\ell+1} +2\a_\ell\, \varphi_\ell \equiv -a_\ell 
\phi_\ell$, where $a_\ell^2 = a_{\ell+1}^2 + 4\a_\ell^2$.  In fact although the 
dilaton coupling constant $a_\ell$ is different in different dimensions 
$\ell$, the related quantity $\Delta$, defined in (\ref{avalue}), is
preserved under dimensional reduction \cite{lpss1}.  The solutions that we
are considering have the feature that the combination of scalar fields
orthogonal to $\phi_\ell$ in $\ell$ dimensions vanishes, \ie $2\a_\ell\,
\phi_{\ell+1} + a_{\ell+1}\, \varphi_{\ell}=0$.  This ensures that a
single-scalar solution in $D$ dimensions remains a single-scalar solution
in all the reduction steps.  Thus we have the following recursive relations 
\bea
&&\fft{\phi_{\ell+1}}{a_{\ell+1}} = \fft{\phi_\ell}{a_\ell} = \cdots =
\fft{\phi_4}{a_4} = 2 (U-\wtd U)\ ,\nonumber\\
&& \varphi_{\ell} = - 2 \a_{\ell} \fft{\phi_\ell}{a_\ell} =
                  -\fft{4}{(\ell-1) (\ell-2)} (U - \wtd U)\ ,
\eea
where $U$ and $\tilde U$ are the functions for the four-dimensional
dilatonic black holes given in section 3.  We find that the metric for the
$(D-4)$-brane in $D$ dimensions is then given by 
\bea
ds^2_{\sst D}\!\!\! &=&\!\!\!e^{-\ft{2(D-4)}{D-2}(U-\wtd U)}\, ds_4^2 +
                e^{\ft{4}{D-2} (U-\wtd U)}\, (dx_1^2 + \cdots + dx_p^2)
\label{dmetric}\\
\!\!\!&=&\!\!\! e^{\ft{4}{D-2} (U-\wtd U)} (-e^{2\wtd U} dt^2 + dx^idx^i)
+ e^{\ft{2(D-4)}{D-2} \wtd U- \ft{4(D-3)}{D-2} U}
\Big(e^{2K} (dr^2 + dz^2) + r^2 d\theta^2\Big)\ ,\nn
\eea
and the dilaton is given by $\phi_{\sst D} = 2a_{\sst D} (U-\wtd U)$.   If
the functions $\wtd U$ and $K$ are those for a Newtonian rod, given by
(\ref{uk}), and the function $U$ is given by (\ref{phians}), the metric
describes a single black $(D-4)$-brane.    The coordinate transformations 
\be
\eta = (\xi - \fft{\hat k^2}{16\xi}) (\cosh\mu)^{\ft{4(D-3)}{\Delta(D-2)}}\ ,
\qquad
x^\mu = \hat x^\mu (\cosh\mu)^{-\ft{4}{\Delta(D-2)}}\ ,
\ee
where $\hat k =(\cosh\mu)^{-4/(\Delta(D-2))}$, put the metric into the
standard isotropic form for a black $(D-4)$-brane, where $c=\tanh \mu$.  The
further transformation $\hat r = (R+ \ft14 \hat k)^2/R$ puts the metric into
the form 
\bea
ds_{\sst D}^2 &=& \Big( 1 + \fft{\hat k}{\hat r} 
\sinh^2\mu\Big)^{-\ft{4}{\Delta(D-2)}}\, (-e^{2f} d\hat t^2 + 
d\hat x^id\hat x^i)\nonumber\\
&& \Big( 1 + \fft{\hat k}{\hat r} \sinh^2\mu \Big)^{\ft{4(D-3)}{\Delta(D-2)}}\,
(e^{-2f} d\hat r^2 + \hat r^2 d\theta^2)\ ,
\eea
where $e^{2f} = 1 -\hat k/\hat r$.  This is the standard form for black 
$(D-4)$-branes discussed in \cite{dlp}.

     Since we again have general solutions given in terms of the harmonic 
function $\wtd U$, we may superpose a set of Newtonian rod potentials, by
taking $\wtd U$ and $K$ to have the forms (\ref{multirod}) and 
(\ref{multisol}).  Equilibrium can again be achieved, without conical 
singularities on the $z$ axis, by taking an infinite line of such rods, 
with equal masses $\ft12 k$, lengths $k$, and spacings $b$.  As discussed 
in section 3, the resulting functions $\wtd U$ and $K$ become 
$z$-independent, and are given by (\ref{zindep}) and (\ref{kzindep}).  Thus 
we can perform a vertical dimensional reduction of the black 
$(D-4)$-brane metric (\ref{dmetric}) in $D$ dimensions to a solution in 
$(D-1)$ dimensions.  For generic values of $k$ and $b$, this solution will 
involve two scalar fields.  However, as discussed in section 3, it will 
describe a single-scalar solution if $k=b$.  In this case, we have $K=\wtd U 
=\log r$, and hence we find that the $(D-1)$-dimensional metric $d\td s_{\sst 
D-1}^2$ , obtained by taking $z$ as the compactification coordinate, so that
$ds^2_{\sst D}= e^{2\a\varphi} d\td s_{\sst D-1}^2 + e^{-2(D-3)\a\varphi} 
dz^2$, is given by
\bea
d\td s_{\sst D-1}^2 &=& - e^{2\wtd U}\, dt^2 + dx^i dx^i + e^{4\wtd U-4 U}
\, dr^2 + r^2 e^{2 \wtd U-4 U}\, d\theta^2 \ ,\nn\\
&=&-r^2 \, dt^2 + dx^i dx^i+ (1-c^2 r^2)^{\ft4{\Delta}}\, 
(dr^2 +d\theta^2)\ .\label{vertox}
\eea
   
    Although the condition that the length $k$ of the rods and their 
spacing $b$ be equal is desirable from the point of view that it gives rise 
to a single-scalar solution in the lower dimension, it is clearly 
undesirable in the sense that the individual single $p$-brane solutions are 
being placed so close together that their horizons are touching.  This 
reflects itself in the fact that the sum over the single-rod potentials is 
just yielding the potential for one rod, of infinite length and infinite 
mass, and accordingly, the higher-dimensional solution just describes a 
single infinitely-massive $p$-brane.  The corresponding vertically-reduced
solution (\ref{vertox}), which one might have expected to describe a black
$((D-1)-3)$-brane in $(D-1)$ dimensions,\footnote{The somewhat clumsy
notation is forced upon us by the lack of a generic $D$-independent name for
a $(D-3)$-brane in $D$ dimensions.} thus does not have an extremal limit. 
This can be understood from another point of view:  A vertically-reduced
extremal solution is in fact a line of uniformly distributed extremal
$p$-branes in one dimension higher.  In order to obtain a black
$((D-1)-3)$-brane in $(D-1)$-dimension that has an extremal limit, we should
be able to take a limit in the higher dimension in which the configuration
becomes a line of extremal $p$-branes. Thus a more appropriate superposition
of black $p$-branes in the higher dimension would be one where the spacing
$b$ between the rods was significantly larger than the lengths of the rods. 
In particular, we should be able to pass to the extremal limit, where the
lengths $k$ tend to zero, while keeping the spacing $b$ fixed.  In this
case, the functions $\wtd U$ and $K$ will take the form (\ref{zindep}) and
(\ref{kzindep}) with $k<b$. Defining $\beta = k/b$, we then find that the
lower-dimensional metric, after compactifying the $z$ coordinate, becomes 
\be
d\hat s_{\sst D-1}^2 = r^{\ft{2\beta(\beta-1)}{D-3}} \, 
\Big( -r^{2\beta}\, dt^2 + dx^i dx^i \Big) + 
r^{-2\beta +\ft{2\beta(\beta-1)}{D-3} } (1-c^2 r^{2\beta} )^{\ft{4}{\Delta}}
\, \Big( r^{2\beta^2}\, dr^2 + r^2\, d\theta^2 \Big) \ .\label{met2}
\ee
This can be interpreted as a black $((D-1)-3)$-brane in $(D-1)$ dimensions.
(In other words, what is normally called a $(D-3)$-brane in $D$ dimensions.)
The extremal limit is obtained by sending $k=b\beta$ to zero and $\mu$ to 
infinity, keeping $b$ and the charge parameter $Q=(\hat k \sinh 
2\mu)/(4\sqrt\Delta)$ finite.  At the same time, we must rescale the $r$ 
coordinate so that $r\rightarrow r (\cosh\mu)^{4/\Delta}$, leading to the 
extremal metric
\be
ds^2 = -dt^2 + dx^i dx^i + \Big( 1- \fft{4\sqrt\Delta Q}{b} \log 
r\Big)^{\ft{4}{\Delta}} (dr^2 + r^2 d\theta^2 )\ .\label{extreme}
\ee
Thus the solution (\ref{met2}) seems to be the natural non-extremal 
generalisation of the extremal $((D-1)-3)$-brane (\ref{extreme}).  Note that
the black solutions (\ref{met2}) involve two scalar fields, as we discussed 
previously, although in the extremal limit the additional scalar decouples.

     In fact the above proposal for the non-extremal generalisation of 
$(D-3)$-branes in $D$ dimensions receives support from a general analysis
of non-extremal $p$-brane solutions.  The usual prescription for
constructing black $p$-branes, involving a single scalar field, as described
for example in \cite{dlp}, breaks down in the case of $(D-3)$-branes in $D$
dimensions, owing to the fact that the transverse space has dimension 2, and
hence $\td d=0$.  Specifically, one can show in general that there is a
universal procedure for ``blackening'' the extremal single-scalar $p$-brane
$ds^2= e^{2A} (-dt^2 + dx^i dx^i) + e^{2B}(dr^2 + r^2 d\Omega^2)$, by
writing \cite{dlp} 
\be
ds^2 = e^{2A}(-e^{2f} dt^2 + dx^i dx^i) + e^{2B}( e^{-2f} dr^2 + r^2 
d\Omega^2) \ ,\label{dlp1}
\ee
where $e^{2f}=1-\hat k r^{-\td d}$, and the functions $A$ and $B$ take the same 
form as in the extremal solution, but with rescaled charges:
\be
e^{-\ft{\Delta(D-2)}{2\td d} A} = e^{\ft{\Delta(D-2)}{2d} B} = 1+ 
\fft{\hat k}{r^{\td d}} \sinh^2\mu\ .\label{dlp2}
\ee
However, the case where $\td d=0$ must be treated separately, and we find
that the black solutions then take the form
\be
ds^2= -(1-\hat k\log r)dt^2 + dx^i dx^i + \fft{1}{r^2}\, \Big(1 + \hat k 
\sinh^2\mu
\log r\Big)^{\ft{4}{\Delta}} \Big((1-\hat k\log r)^{-1}\, dr^2 + 
d\theta^2 \Big) 
\ .\label{dlp3}
\ee
In the extremal limit, \ie  $\hat k\rightarrow 0$ and
$\mu\rightarrow\infty$, the metric becomes 
\be
ds^2= -dt^2 + dx^i dx^i + (1 + Q R)^{\ft{4}{\Delta}} (dR^2 + d\theta^2) \ ,
\label{ext3}
\ee
where $R=\log r$.  Unlike the situation for non-zero values of $\td d$, 
where the analogous limit of the black $p$-branes gives a normal isotropic 
extremal $p$-brane, in this $\td d=0$ case the extremal limit describes a 
line of $(D-3)$-branes in $D$ dimensions, lying along the $\theta$ 
direction, rather than a single $(D-3)$-brane.  (In fact this line of 
$(D-3)$-branes can be further reduced, by compactifying the $\theta$ 
coordinate, to give a domain-wall solution in one lower dimension 
\cite{clpst,bdgpt}.) Thus it seems that there is no appropriate 
single-scalar non-extremal generalisation of an extremal $(D-3)$-brane in 
$D$ dimensions, and the two-scalar solution (\ref{met2}) that we obtained by
vertical reduction of a black $(D-3)$-brane in one higher dimension is the 
natural non-extremal generalisation.

\section{Conclusions}

      In this paper, we raised the question as to whether one can generalise 
the procedure of vertical dimensional reduction to the case of non-extremal 
$p$-branes.  It is of interest to do this, since, combined with the more 
straightforward procedure of diagonal dimensional reduction, it would provide 
a powerful way of relating the multitude of black $p$-brane solutions of 
toroidally-compactified M-theory, analogous to the already well-established 
procedures for extremal $p$-branes.  Vertical dimensional reduction 
involves compactifying one of the directions transverse to the $p$-brane 
world-volume. In order to achieve the necessary translational invariance 
along this direction, one needs to construct multi-center $p$-brane
solutions in the higher dimension, which allow a periodic array of
single-center solutions to be superposed.  This is straightforward for 
extremal $p$-branes, since the no-force condition permits the construction 
of arbitrary multi-center configurations that remain in neutral equilibrium. 
No analogous well-behaved multi-center solutions exist in general in the 
non-extremal case, since there will be net forces between the various 
$p$-branes.  However, an infinite periodic array along a line will still be 
in equilibrium, albeit an unstable one.  This is sufficient for the 
purposes of vertical dimensional reduction.

    The equations of motion for general axially-symmetric $p$-brane
configurations are rather complicated, and in this paper we concentrated on
the simpler case where the transverse space is 3-dimensional.  This leads to
simplifications in the equations of motion, and we were able to obtain the
general axially-symmetric solutions for charged dilatonic non-extremal
$(D-4)$-branes in $D$ dimensions.  These solutions are determined by a
single function $\wtd U$ that satisfies a linear equation, namely the
Laplace equation on a flat cylindrically-symmetric 3-space, and thus
multi-center solutions can be constructed as superpositions of basic
single-center solutions.  The single-center $p$-brane solutions correspond
to the case where $\wtd U$ is the Newtonian potential for a rod of mass $k$
and length $\ft12 k$. 

    The rather special features that allowed us to construct general 
multi-center black solutions when the transverse space is 3-dimensional also
have a counterpart in special features of the lower-dimensional solutions
that we could obtain from them by vertical dimensional reduction.  The
reduced solutions are expected to describe non-extremal $(D-3)$-branes in
$D$ dimensions.  Although a general prescription for constructing
single-scalar black $p$-branes from extremal ones for arbitrary $p$ and $D$
was given in \cite{dlp}, we found that an exceptional case arises when
$p=D-3$.  In this case the general analysis in \cite{dlp} degenerates, and
the single-scalar black solutions take the form (\ref{dlp3}), rather than
the naive $\td d\rightarrow 0$ limit of (\ref{dlp1}) and (\ref{dlp2}) where
one would simply replace $r^{-\td d}$ by $\log r$. The extremal limit of
(\ref{dlp3}) in fact fails to give the expected extremal $(D-3)$-brane, but
instead gives the solution (\ref{ext3}), which describes a line of
$(D-3)$-branes.  Interestingly enough, we found that the vertical reduction
of the non-extremal $p$-branes obtained in this paper gives a class of
$(D-3)$-branes which are much more natural non-extremal generalisations of
extremal $(D-3)$-branes.  In particular, their non-extremal limits {\it do}
reduce to the standard extremal $(D-3)$-branes.  The price that one pays for
this, however, is that the non-extremal solutions involve two scalar fields
(\ie the original dilaton of the higher dimension and also the Kaluza-Klein
scalar), rather than just one linear combination of them.  Thus we see that
a number of special features arise in the cases we have considered.  It
would be interesting to see what happens in the more generic situation when
$\td d>0$.

\section*{Acknowledgement}

      We are grateful to Gary Gibbons for useful discussions, and to Tuan
Tran for drawing our attention to some errors in an earlier version of the
paper.  H.L.\ and C.N.P.\ are grateful to SISSA for hospitality in the early
stages of this work. K.-W.X.\ is grateful to TAMU for hospitality in the
late stages of this work.

\end{document}